\documentclass[12pt,preprint]{aastex}

 
\begin{document}

\title{Temperature Bias in Measurements of The Hubble Constant Using The Sunyaev-Zeldovich Effect}
\author{Wen-Ching Lin\altaffilmark{1}, Michael L. Norman\altaffilmark{2}, Greg L. Bryan\altaffilmark{3}}
\altaffiltext{1}{Department of Astronomy, University of Illinois at Champaign Urbana, Urbana, IL 61801, U.S.A.}
\altaffiltext{2}{CASS, University of California at San Diego, La Jolla, CA 92093, U.S.A.}
\altaffiltext{3}{The Denys Wilkinson Building, Keble Road, Oxford, OX1 3RH, UK}

\begin{abstract}

Measurements of the Hubble constant to distant galaxy clusters using the Sunyaev-Zeldovich effect are systematically low in comparison to values obtained by other means. These measurements usually assume a spherical isothermal $\beta$ model for the intracluster medium (ICM). Formation processes and recent mergers guarantee that real clusters are neither spherical nor isothermal. We present the results of a statistical analysis of {\em temperature bias} in $H_0$ determinations in a sample of 27 numerically simulated X-ray clusters drawn from a $\Lambda$CDM model at z=0.5 (the sample is online at http://sca.ncsa.uiuc.edu). We employ adaptive mesh refinement to provide high resolution (15.6 h$^{-1} kpc$) in cluster cores which dominate the X-ray and radio signals (sample images on line at http://cosmos.ucsd.edu/$\sim$wenlin/SZ/sz$\_$cluster.html). The clusters possess a variety of shapes and merger states which are computed self-consistently in a cosmological framework assuming adiabatic gas dynamics. We derive the value of $H_0$ by computing the angular diameter distance to each cluster along three orthogonal lines of sight. Fitting synthetic X-ray and y-parameter maps to the standard isothermal beta model, we find a broad, skewed distribution in $f\equiv H_0(SZ)/H_0(actual)$ with a mean, median, and standard deviation of 0.89, 0.83 and 0.32 respectively, where $H_0(SZ)$ is the value of $H_0$ derived by using Sunyaev-Zeldovich effect method and $H_0(actual)$ is the value used in the cosmological simulation. We find that the clusters' declining temperature profiles systematically lower estimates of $H_0$ by 10\% to 20\%. The declining temperature profile of our adiabatic system is consistent with the result including radiative cooling and supernovae feedback (Loken et al. 2002). We thereby introduce a non-isothermal $\beta$ model as an improvement. Applying the non-isothermal $\beta$ model to the refined sample with well-fitted temperature profile, the value of f improves 9$\%$ relative to the actual value. The study of the morphology and the clumping effects conclude that these two factors combine to induce scatter in f of $\pm$ 30 $\%$.


\end{abstract}

\section{INTRODUCTION} 

    The Sunyaev-Zeldovich effect (SZE) is a spectral distortion of the cosmic microwave background (CMB) due to the inverse Compton scattering of CMB photons by high energy electrons in the hot intracluster medium (ICM) (Sunyaev $\&$ Zeldovich 1970,1972). At low frequencies ($\lesssim$ 218 GHz) the distortion appears to be a decrement and thereby the peak of CMB spectrum ($\sim$ 160 GHz) intensity is reduced. This effect has been suggested as an important tool for studying cosmology and astrophyics (Sunyaev $\&$ Zeldovich 1980; Birkinshaw 1999 for more references). One important application of observing the SZE of galaxy clusters is the determination of the Hubble constant ($H_0$). By combining SZE data with X-ray data from a cluster of galaxies, one can measure the distance to the cluster and thereby derive the value of $H_0$ under a given cosmological model.


  A list of estimates of $H_0$ based on this method are collected by Birkinshaw (1999). There are also some new results with better data and analysis (eg. Reese et al. 2000; Mason et al. 2001; Pointecouteau et al. 2002). The value of $H_0$ from these findings are systematically lower than by other methods (Table.~\ref{tbl:H_0.tbl}). This systematic discrepancy is due to the lack of information about the intracluster gasdynamics, which forces observers to assume that clusters are spherically symmetric, isothermal and in hydrostatic equilibrium. One well-known model based on these assumptions and frequently used by observers is the isothermal $\beta$ model (Cavaliere \& Fusca-Femiano 1976, 1978). However, drastic improvements of the X-ray observations in the past decade have revealed the deficiencies of this model. One particular important fact that contradicts the isothermal $\beta$ model is the presence of clumping and substructures in clusters which characterize a cluster merger. (eg. Jones $\&$ Forman 1984; Gomez et al. 1997; Bilton et al. 1998; Neumann et al. 1999; van Dokkum et al. 1999). Moreover, non-isothermal temperature variations and decreasing temperature profiles have begun to be observed (Roettiger et al. 1995; Markevitch et al. 1997; Wang 1998; Honda et al. 1998).

  To investigate the consquences of the isothermal $\beta$ model for $H_0$, we perform a statistical analysis on 27 rich clusters drawn from a fully hydrodynamic simulation with 15.6/h kpc resolution at cluster cores. We consider the influence of each assumption but focus on the temperature structures of clusters. In this paper, we describe the theory of the isothermal $\beta$ model and the calculation of $H_0$ in $\S 2$ and numerical methods of simulations in $\S 3$. We first derive the values of $H_{0}$ by applying the isothermal $\beta$ model in $\S 4.1$. Some statistics of the $\beta$ parameter are discussed in $\S 4.2$; the non-isothermality and asphericity of the model are elucidated in $\S 4.3$ and $\S 4.4$. We then introduce a non-isothermal $\beta$ model to improve the analysis method in $\S 5$. We also include the study of cluster morphology and the clumping effect in $\S 6$ and $\S 7$ and conclude in $\S 8$.

\section{ISOTHERMAL $\beta$ MODEL :  SZE AND $H_{0}$}
 It is convenient to introduce a parametrized model to account the scattered photons in the intracluster gas. One popular model is the isothermal $\beta$ model, which  assumes that the electron temperature $T_{e}$ is constant and the electron number density has a spherically symmetric distribution:
\begin{equation}
   n_{e}(r)=n_{e0} (1+ \frac{r^{2}}{r^{2}_{c}})^{-\frac{3\beta}{2}}
\label{eqn:ne.eqn}
\end{equation}
 where $r_{c}$ is the core radius of the cluster, $n_{e0}$ is the central electron number density, and $\beta$ is a parameter close to $\frac{2}{3}$. (Cavaliere \& Fusco-Femiano 1976, 1978). Over decades, this model has been much used in fitting the structures of X-ray clusters (Savazin 1988). If the electron density in a cluster atmosphere is $n_{e}$(r), the Comptonization parameter along a particular line of sight is given by : 
\begin{eqnarray}
   y & = & \int \sigma_{T} n_{e}(r) \frac{k_{B} T_{e}}{m_{e} c^{2}} dl \nonumber \\
               & = & D_A \int \sigma_{T} n_{e}(r) \frac{k_{B} T_{e}}{m_{e} c^{2}} d\xi
\label{eqn:y.eqn}
\end{eqnarray}
  where $k_{B}$ is the Boltzmann constant, $\sigma_{T}$ is the Thompson cross section, $m_{e}$ is the electron mass, and c is the speed of light; the line of sight \textit{l} is given by $D_A$ $\xi$, where $D_A$ is the angular diameter distance to the cluster and $\xi$ is the corresponding renormalized parameter.

 It can be shown that the X-ray surface brightness is given by :
\begin{eqnarray}
   S_{x} & = & \frac{1}{4 \pi(1+z)^{4}}\int n_{e}(r) n_{H}(r) \Lambda_{eH} dl \nonumber \\
               & = & \frac{1}{4 \pi(1+z)^{4}} D_A \int n_{e}(r) n_{H}(r) \Lambda_{eH} d \xi
\label{eqn:Sx.eqn}
\end{eqnarray}

   where $S_{x}$ is the bolometric X-ray surface brightness, $\textit{z}$ is the redshift of the cluster, $n_{H}$ and $n_{e}$ are the hydrogen and electron number densities, and $\Lambda_{eH}$ is the X-ray cooling function. The maxima of the X-ray surface brightness and SZ y-parameter maps are denoted by $S_{x0}$ and $y_0$. By inserting Eqn.(\ref{eqn:ne.eqn}) to Eqn.(\ref{eqn:y.eqn}) and Eqn.(\ref{eqn:Sx.eqn}), we can eliminate $n_{e0}$ from Eqn.(\ref{eqn:y.eqn}) and Eqn.(\ref{eqn:Sx.eqn}) and thereby solve for the angular diameter distance :
\begin{eqnarray}
   D_A & = & \frac{y_{o}^{2}}{4 \pi (1+z)^{4} S_{xo}} (\frac{m_{e} c^{2}}{k_{B} T_{e}} )^{2} \frac{ \frac{\mu_{e}}{\mu_{H}} \Lambda_{eH}}{ \pi^{\frac{1}{2}} \sigma_{T}^{2}}  \frac{\Gamma(3 \beta -0.5)}{\Gamma(3 \beta)} (\frac{\Gamma(\frac{3}{2} \beta )}{\Gamma( \frac{3}{2} \beta - 0.5 )})^{2} \frac{1}{\theta_{c}}
\label{eqn:Da.eqn}
\end{eqnarray}
 where $\Gamma(x)$ is the Gamma function and $\theta_c$ is the core size of the cluster. Since the angular diameter distance  $D_A$ is a function of $H_0$ and the cosmology, $H_{0}$ can be calculated under a given cosmological model (Birkinshaw 1999; Reese 2000).

\section{NUMERICAL METHOD}
\subsection{Cosmological Simulation}

 The simulated clusters used in this project are part of our numerical ``catalog'' of clusters which are available at http://sca.ncsa.uiuc.edu (Simulated Cluster Archive; Norman et al. 1999). The cosmological simulation is performed on the supercomputer Origin2000 at NCSA using our adaptive mesh refinement (AMR) cosmological hydrodynamic code ENZO. ENZO incorperates a Lagrangian particle-mesh (PM) algorithm to follow the collisionless dark matter. The equations of gas dynamics are calculated by the piecewise parabolic method (PPM; Collela $\&$ Woodward, 1984) which yields high fidelity in the ICM temperature and density distributions. These numerical techniques have been described in more detail previously (Bryan et al. 1995; Bryan et al. 1999; Norman $\&$ Bryan, 1999).

 The cosmological parameters used in this simulation are: $\Omega_0$ = 0.3, $\Omega_b$ = 0.026, $\Omega_{\Lambda}$ = 0.7, h = 0.7 and $\sigma_8$ = 0.928. The high resolution adaptively refined simulations are performed after an initial low-resolution simulation is used to identify clusters. The simulation starts in an $128^{3}$ base grid (level 0 grid; L0), 256/h Mpc box at redshift 30. We then put two subgrids (L1 $\&$ L2) covering each cluster to refine the Lagrangian volume there. Within the L2 grid, up to five more levels (L3 $\to$ L7) of refinements are automatically introduced during the simulation. From L1, each grid is twice the resoultion of the previous one and therefore the best spatial resolution is 15.6/h kpc. (Loken et al. 2002)


\subsection{Analysis Method}
   The simulated X-ray clusters are all at redshift 0.5. We observe these clusters along three lines of sight parallel to the edges of the box. Then we create synthetic maps of bolometric X-ray surface brightness and SZ y-parameter by projecting the two dimensional maps along the observational lines of sight. The projected surface brightness and the y parameter can be written as:
\begin{equation}
  S_x(\theta) \equiv S_{x0}(1+(\frac{\theta}{\theta_c})^2)^{\frac{1}{2}-3\beta}
\label{eqn:Sx_profile.eqn}
\end{equation}
\begin{equation}
  y(\theta) \equiv y_0(1+(\frac{\theta}{\theta_c})^2)^{\frac{1}{2}-\frac{3\beta}{2}}
\label{eqn:y_profile.eqn}
\end{equation}
 where $\theta$ is the angular size of the cluster in the sky, and $\theta_c$ is the corresponding core size; $S_{x0}$ and $y_0$ are peaks in the projected maps. We derive both profiles by calculating the annular average of the projected surface brightness and y-parameter maps. Applying a three-parameter $\chi^{2}$ fitting on the profiles to Eqn.(\ref{eqn:Sx_profile.eqn}) and Eqn.(\ref{eqn:y_profile.eqn}), we can determine the central values ($S_{x0}$ or $y_0$) as well as $\theta_c$ and $\beta$. The $\beta$ value and $\theta_c$ used in Eqn.(\ref{eqn:Da.eqn}) are from fitting the X-ray surface brightness since the X-ray signal is much stronger than the SZ signal and dominate the final result. The X-ray cooling function $\Lambda_{eH}$ is calculated by a Raymond-Smith code with metalicity 0.3 solar.

\section{RESULTS OF THE ISOTHERMAL $\beta$ MODEL}

\subsection{Bias Due to the Isothermal Assumption} 
  We analyze 27 numerically simulated clusters using the isothermal $\beta$ model and thereby derive the value of $H_0$. Applying the analysis procedure mentioned in $\S 3.2$, we compute the values of $H_0$ along three orthogonal lines of sight: x, y, and z. The resulting values of $H_0$ are normalized by 70, which is the value adopted in the cosmological simulation. We denote the normalized Hubble constant by f and find a broad, skewed distribution in f with a mean and median of 0.897 and 0.829, respectively. The standard deviation is 0.32 if we parametrize the f distribution by a Gaussian (Table.~\ref{tbl:f_dis_ori.tbl}).  In order to identify the origin of this bias in $H_0$, we first test the assumption of ``isothermality''. This is done by setting the temperature of the entire cluster equal to the emission weighted temperature and creating the corresponding bolometric X-ray and the SZE y-parameter maps. The results of analyzing these 27 isothermal clusters are displayed in Table.~\ref{tbl:f_dis_iso.tbl}, where the mean and median are 1.111 and 1.093 while the standard deviation is 0.297. Comparing the results from these two groups of clusters, we conclude that the applying an ``isothermal'' assumption on a non-isothermal cluster lowers the value of $H_0$ by around 20 $\%$ compared to a real isothermal cluster. It is also noticeable that the standard deviation of the f distribution is close in both cases, which implies that the broad scattering in f is not mainly caused by the non-isothermality of clusters.

\subsection{ $\beta$ Value Distribution}
\begin{figure}
\plotone{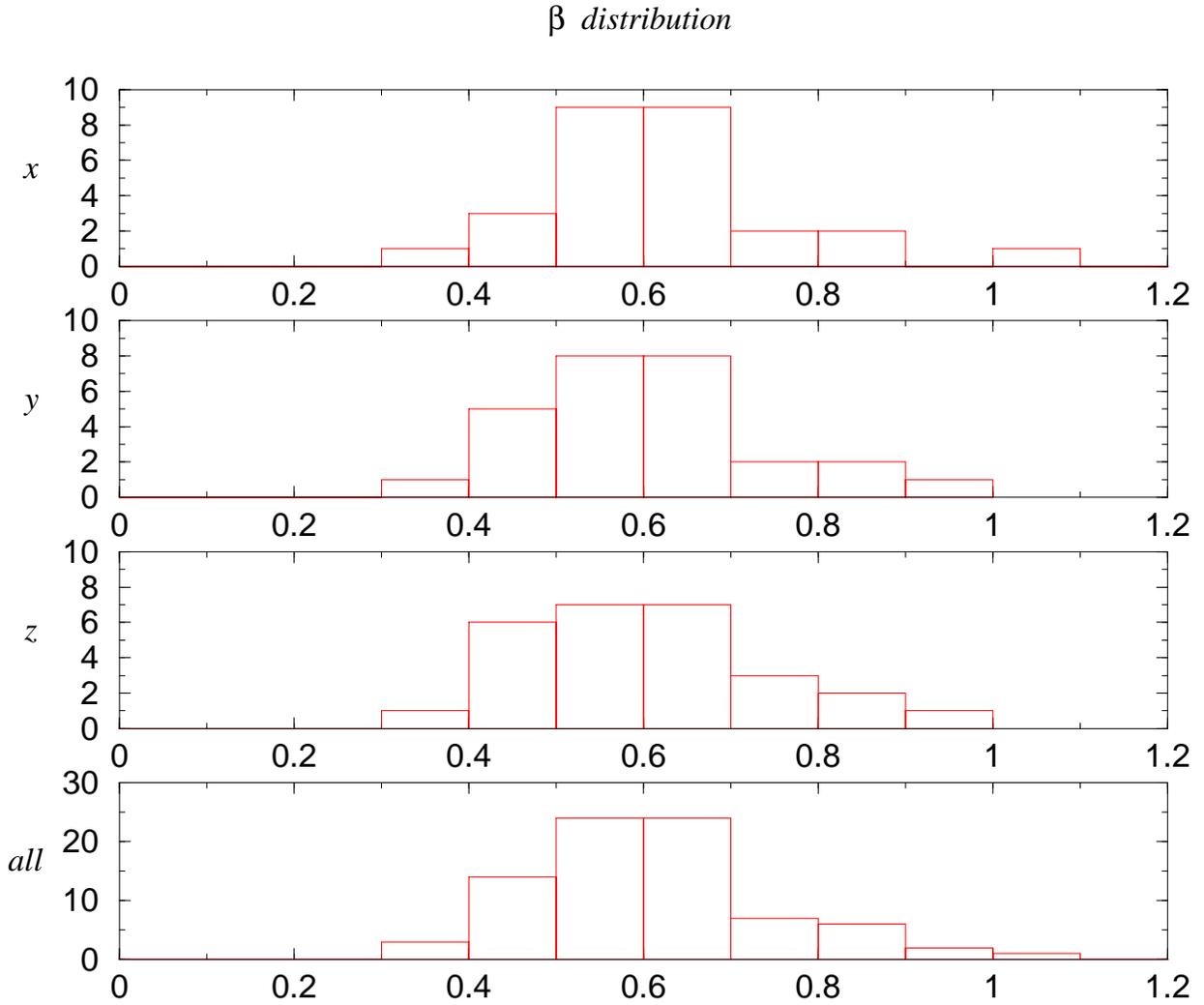}
\caption{\label{fig:beta_all.fig} The distribution of $\beta$ values derived from fitting the X-ray surface brightness. The top three figures show the results along the x, y, and z observational lines of sight and the bottom one combines the three lines of sight data.}
\end{figure}

\begin{figure}
\plotone{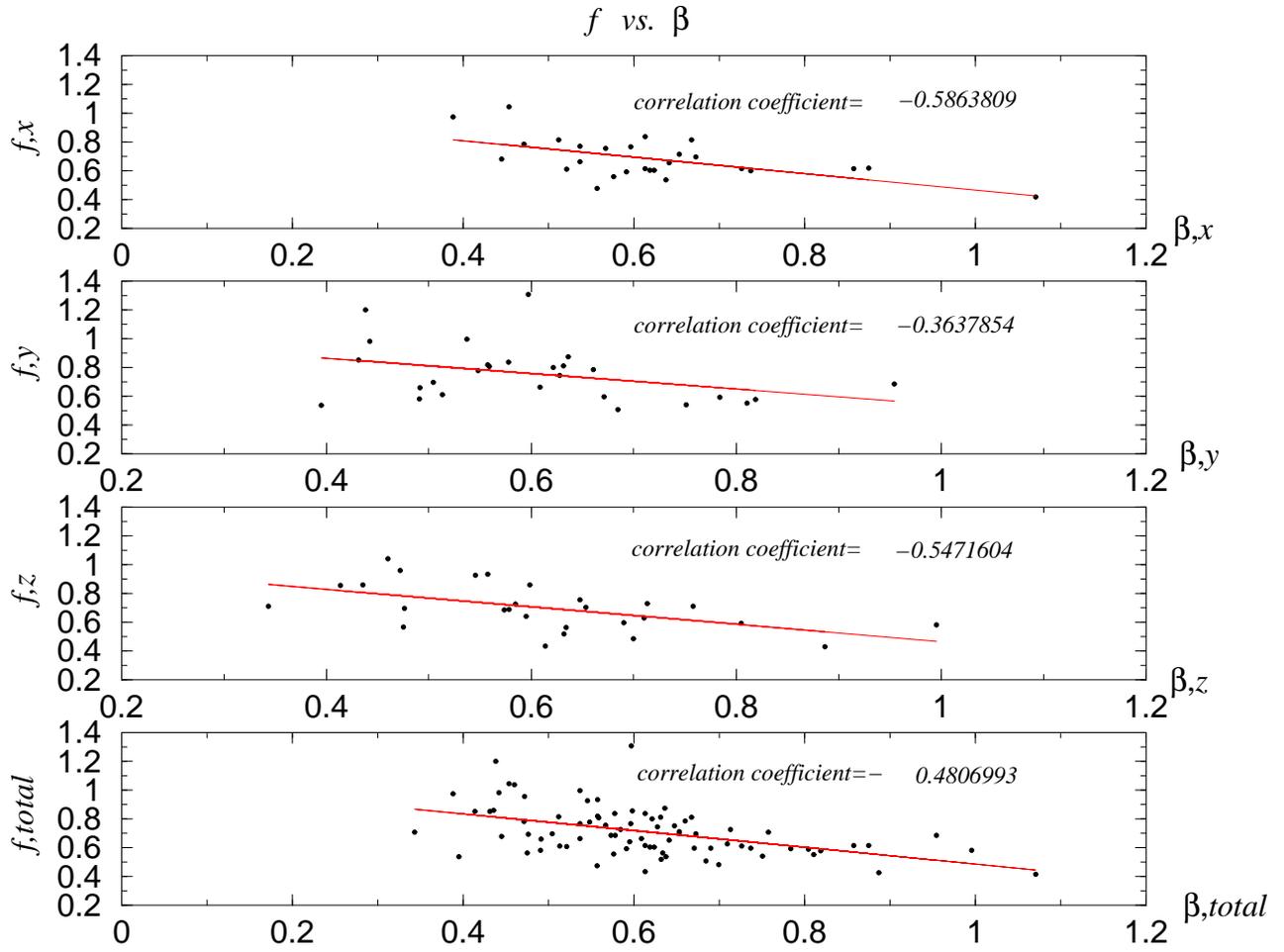}
\caption{\label{fig:f_vs_beta.fig} The correlation between the normalized Hubble constant f and the value $\beta$. The top three figures display the results of the original 27 clusters along x, y, and z directions, respectively. The bottom one shows the result by combining the three lines of sight data.}
\end{figure}

\begin{figure}
\plotone{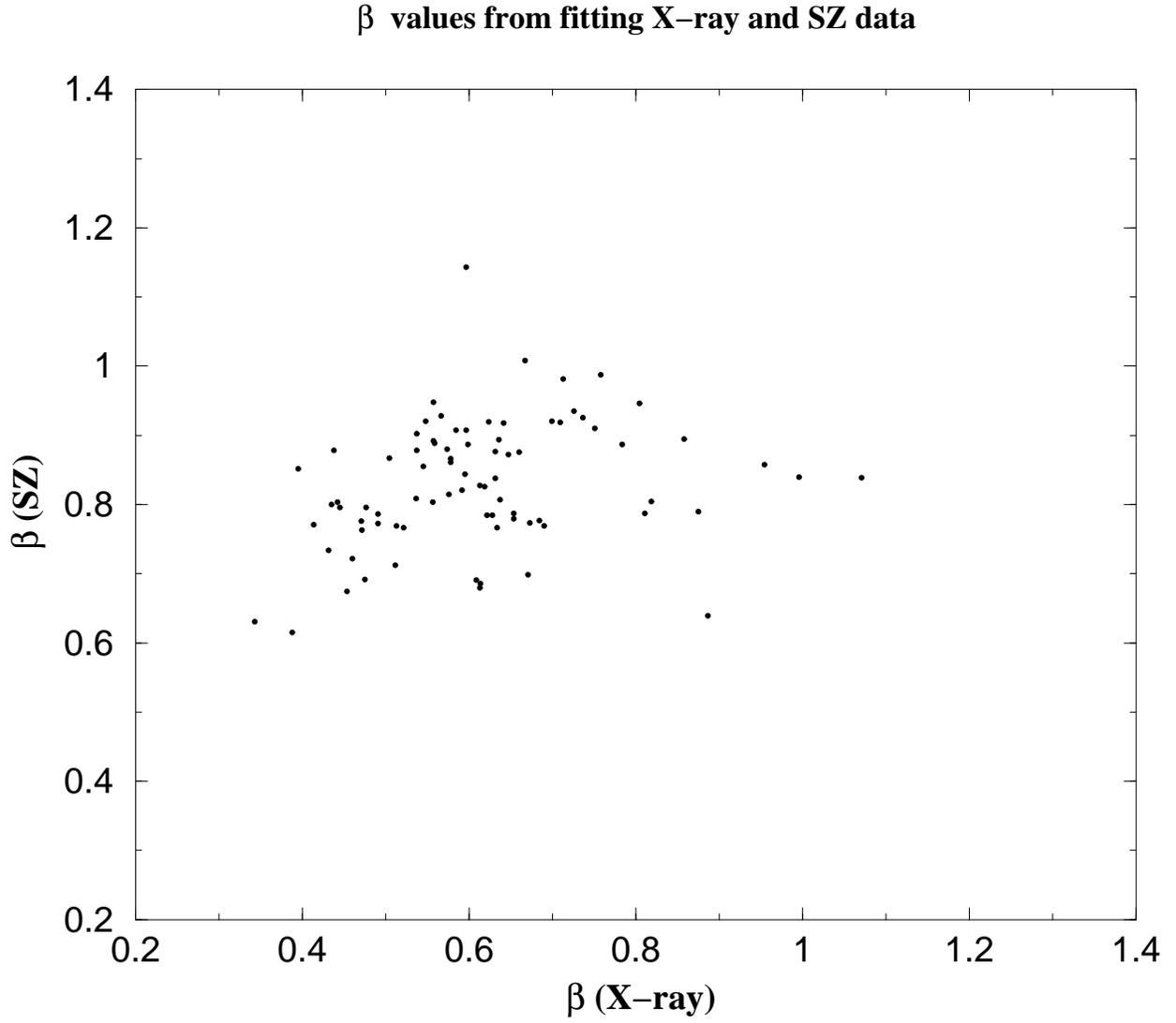}
\caption{\label{fig:beta_xray_sz.fig} $\beta$(X-ray) is the value of $\beta$ derived from fitting the X-ray surface brightness. $\beta$(SZ) is the value of $\beta$ derived from the SZE y-parameter contour map.}

\end{figure}

  Based on fitting the X-ray surface brightness maps, some previous simulations have shown that the mean value of $\beta$ is around 0.8 (Metzler $\&$ Evrard, 1994) while the observational values are approximately 0.6 (eg. Mauskope et. al., 2000). The mean value of $ \beta$ in our simulation are 0.6210, 0.6054, and 0.6130 along 3 distinct lines of sight and is 0.61 if we combine the 3 line-of-sight results (Fig.~\ref{fig:beta_all.fig}; Table.~\ref{tbl:beta.tbl}). This indicates that our cosmological hydrodynamic code produces more realistic results than other codes (P3MSPH code, Metzler $\&$ Evrard, 1994). Another interesting result here is that we actually find a correlation between f and $\beta$ (Fig.~\ref{fig:f_vs_beta.fig}), which deserves more investigation. We also fit the $\beta$ parameter from the SZ contour map as well as from the X-ray surface brightness map. The $\beta$ value derived from the SZ data is typically higher than the one derived from X-ray data. Fig.~\ref{fig:beta_xray_sz.fig} shows the relation of $\beta$ values derived from the two different methods.

\subsection{Statistics of the Non-isothermality}
\begin{figure}
\plotone{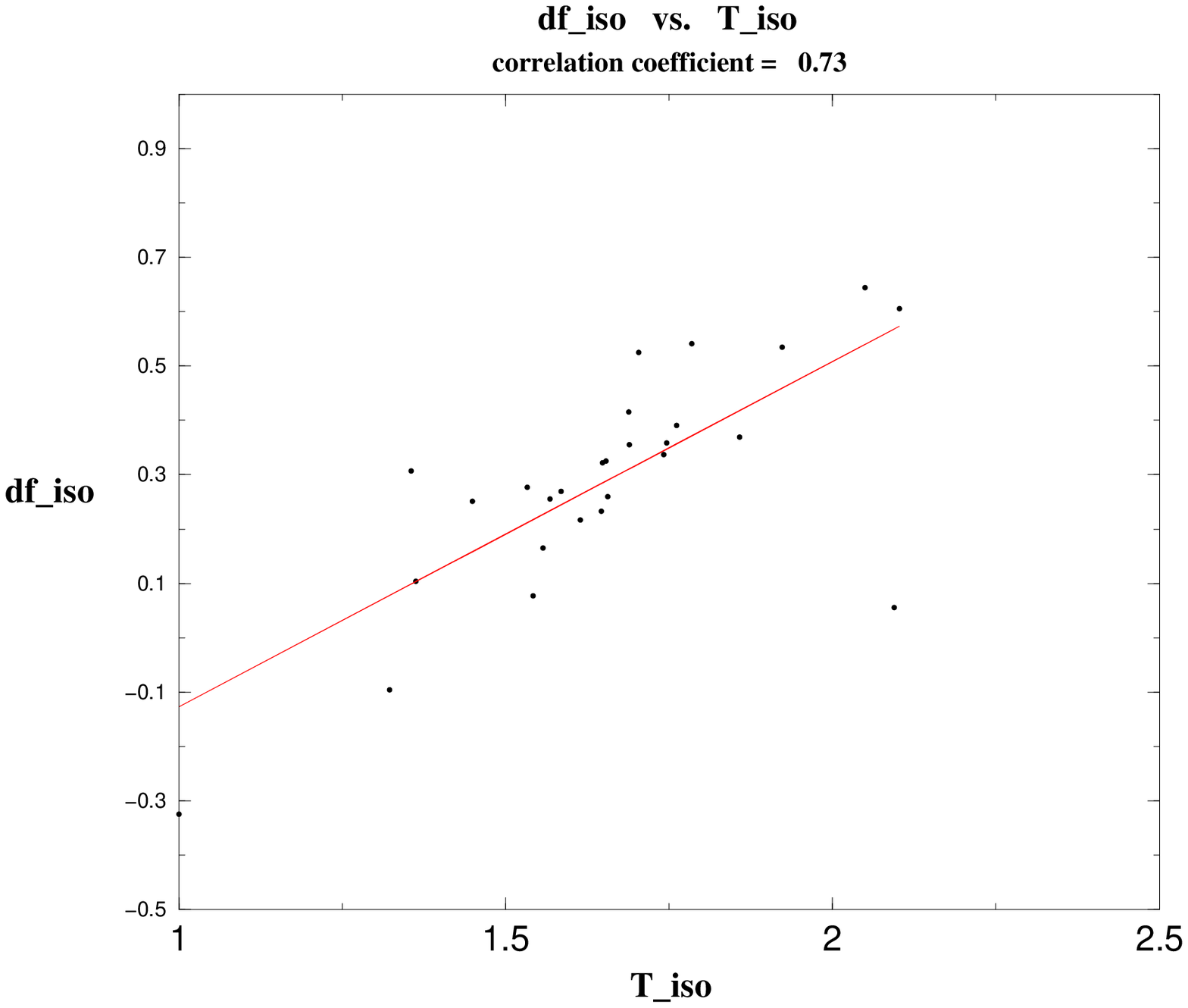}
\caption{\label{fig:df_vs_Tiso.fig} Bias of f due to the non-isothermality. $df_{iso}$ is defined as $\frac{f(iso)-f(ori)}{f(ori)}$, where f(iso) is the result of the 27 isothermal clusters while f(ori) is the result from the original ones. $T_{iso}$ denotes the degree of non-isothermality and is defined as $\frac{T_0}{<T_x>}$.}
\end{figure}

  A rich cluster possesses a high potential well at the central region where most dark matter concentrates. At hydrostatic equalibrium, the temperature follows the gravitational potential. Since the potential decreases from the central core to the outer region, a decreasing temperature profile is expected ($\S 4.5$). We define $T_{iso}$ as $\frac{T_0}{<T_x>} $ to describe the ``non-isothermality'' of a cluster, where $T_0$ and $<T_x>$ are the central and emission-weighted temperatures, respectively. Thus $T_{iso}$ equals 1 when a cluster is exactly isothermal. We denote the bias of the normalized Hubble constant f as $df_{iso}$, which is defined as $\frac{f(iso)-f(ori)}{f(ori)}$, where $\textit{f(iso)}$ and $\textit{f(ori)}$ are the normalized Hubble constant averaged over three lines of sight for the isothermal and the original clusters, respectively ($\S 4.1$). The strong correlation (correlation coefficient equals 0.73) between these two variables confirms that the huge bias is due to the ``isothermal'' assumption (see Fig.~\ref{fig:df_vs_Tiso.fig}). 


\subsection{Elongation and $H_0$}
\begin{figure}
\plotone{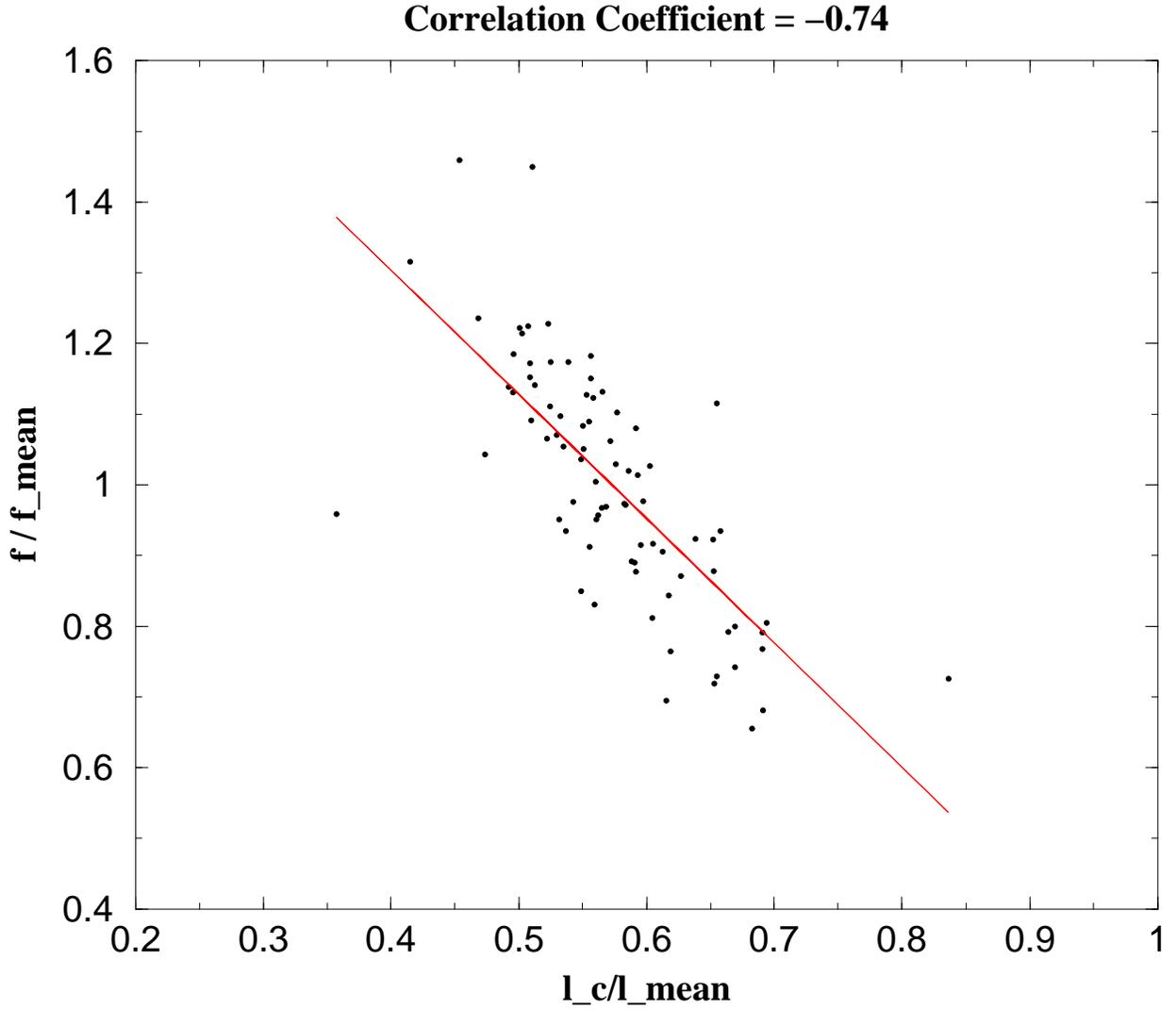}
\caption{\label{fig:lc_f.fig} For each cluster, $\frac{f}{f_{mean}}$ represents the bias of f in each observational direction relative to the $f_{mean}$ which is the averaged value along the three observational lines of sight for that given cluster. Similarly,  $\frac{l_c}{l_{mean}}$ denotes the bias of the characteristic length along one observational line of sight relative to $l_{mean}$ which is the characteristic size of the given cluster.} 
\end{figure}

   To study the elongation effect on $H_0$, we define a characteristic length for a cluster at certain direction as:
\begin{equation}
   l_{c} \equiv \frac{ (\int \rho dl)^{2}}{\int \rho^{2} dl}
\end{equation}
  where $\textit{dl}$ integrates through the peak of a projected X-ray surface brightness map along the observational line of sight. The results of three different lines of sight are denoted by $l_{c,x}$, $l_{c,y}$, and $l_{c,z}$. Since the size of a cluster varies, we normalize $l_{c,x}$, $l_{c,y}$, and $l_{c,z}$ by the value $l_{mean}$ = $\sqrt{l_{c,x}^2 + l_{c,y}^2 + l_{c,z}^2} $. On the other hand, for each single cluster, the normalized Hubble constant along three observational lines of sight $f_x$, $f_y$, and $f_z$ are renormalized by their mean value $f_{mean}$ = $\frac{1}{3}$ ($f_x$+$f_y$+$f_z$). Because $f_{mean}$ varies in different clusters due to different temperature structures ($\S 4.3$), we reduce corresponding effects by renormalizing f by $f_{mean}$ . Fig.~\ref{fig:lc_f.fig} shows the strong negtive correlation (correlation coefficient equals -0.74) between $\frac{l_c}{l_{mean}}$ and $\frac{f}{f_{mean}}$, which indicates that for a given aspherical cluster, the observed value of $H_0$ is smaller along the longer axis. This result is consistent with the study by Roettiger et al. (1997). 

\section{IMPROVED METHOD : NON-ISOTHERMAL $\beta$ MODEL}
\begin{figure}
\plotone{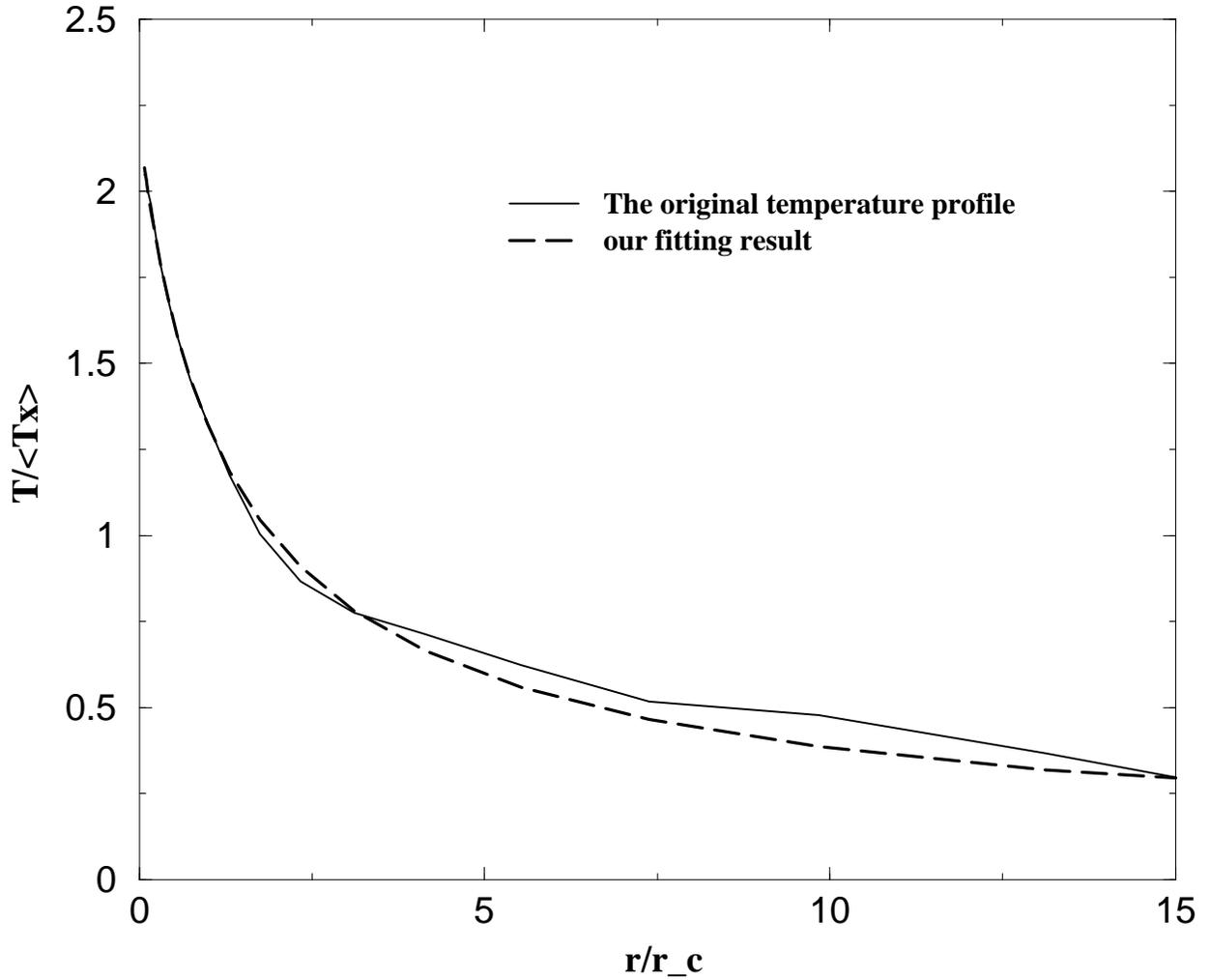}
\caption{\label{fig:cl0022_sample_p.fig} The temperature profile of a sample cluster. The black-solid curve denotes the temperature profile of the cluster cl0022; the red-dotted curve is the fitting result for the profile. The radial distance is normalized by the core radius of the cluster and the temperature is normalized by the emission weighted temperature.}
\end{figure}

   Our study of  the 27 numerically simulated clusters reveals an universal temperature profile. The best fitting formula for the temperature profile is :
\begin{equation}
   T(r)=T_{0} (1+ \frac{r}{r_{c}})^{-\alpha}
\label{eqn:tem_profile.eqn}
\end{equation}
   where $T_0$ is the central temperature of a cluster and $r_c$ is the core radius. At high redshift, such as 0.5 in our case, a lot of clusters are still dynamically young and non-relaxed, which results in the presence of some temperature clumps inside a cluster and thereby an irregular temperature profile. Here we disregard this type of clusters and concentrate on clusters without temperature clumps. Our refined sample includes 16 clusters with well-fitted temperature profiles. The average value of $\alpha$ drawn from these 16 clusters is 0.56. Fig.~\ref{fig:cl0022_sample_p.fig} shows a sample cluster fitted by this formula.

   Based on the dramatic improvement of X-ray data from Chandra and XMM, we anticipate getting temperature profiles of clusters in the near future. However, if the temperature profile is not available, one can still use non-isothermal $\beta$ model by taking the temperature profile derived from our simulation. If the temperature profile of a cluster follows Eqn.(\ref{eqn:tem_profile.eqn}), the central temperature $T_0$ and emission-weighted temperature $<T_x>$ are linked by :
\begin{equation}
  <T_x> = \frac{\int_0^R T(r) e(r) 4 \pi r^2 dr}{\int_0^R e(r) 4 \pi r^2 dr}
\end{equation}
  where R is the radial size of cluster and normally chosen to be the virial radius, which is around 1 Mpc for rich clusters; e(r) $\propto$ $n_e$(r) $n_H$(r) $\Lambda$(r), where the bolometric emissivity $\Lambda$(r) $\propto$ $T(r)^{1/2}$, $n_e$(r) and $n_H$(r) are the number density profiles of electrons and the protons. Applying this temperature fitting formula to the final 16 clusters, we improve the mean value of $H_0$ by 9 $\%$ (Table.~\ref{tbl:Bmodel_compare.tbl}). The lack of clumpings in the temperature profiles indicates the relaxation of the clusters, which results in the refined sample having less scattering in the f distribution than the original ones.
 
 One might argue that this decrease in the temperature profile is a result of the insufficient physics put in the simulation of the adiabatic clusters which do not have any cooling. However, some recent simulations including radiative cooling and supernovae feedback also show declining temperature profiles. These temperature profiles actually drop even more steeply than the adiabatic case (Loken et al. 2002) and therefore the improvement benefits from the non-isothermal $\beta$ model would be even more for the cooling clusters.

\section{CLUSTER MORPHOLOGY AND $H_{0}$}
\begin{figure}
\plotone{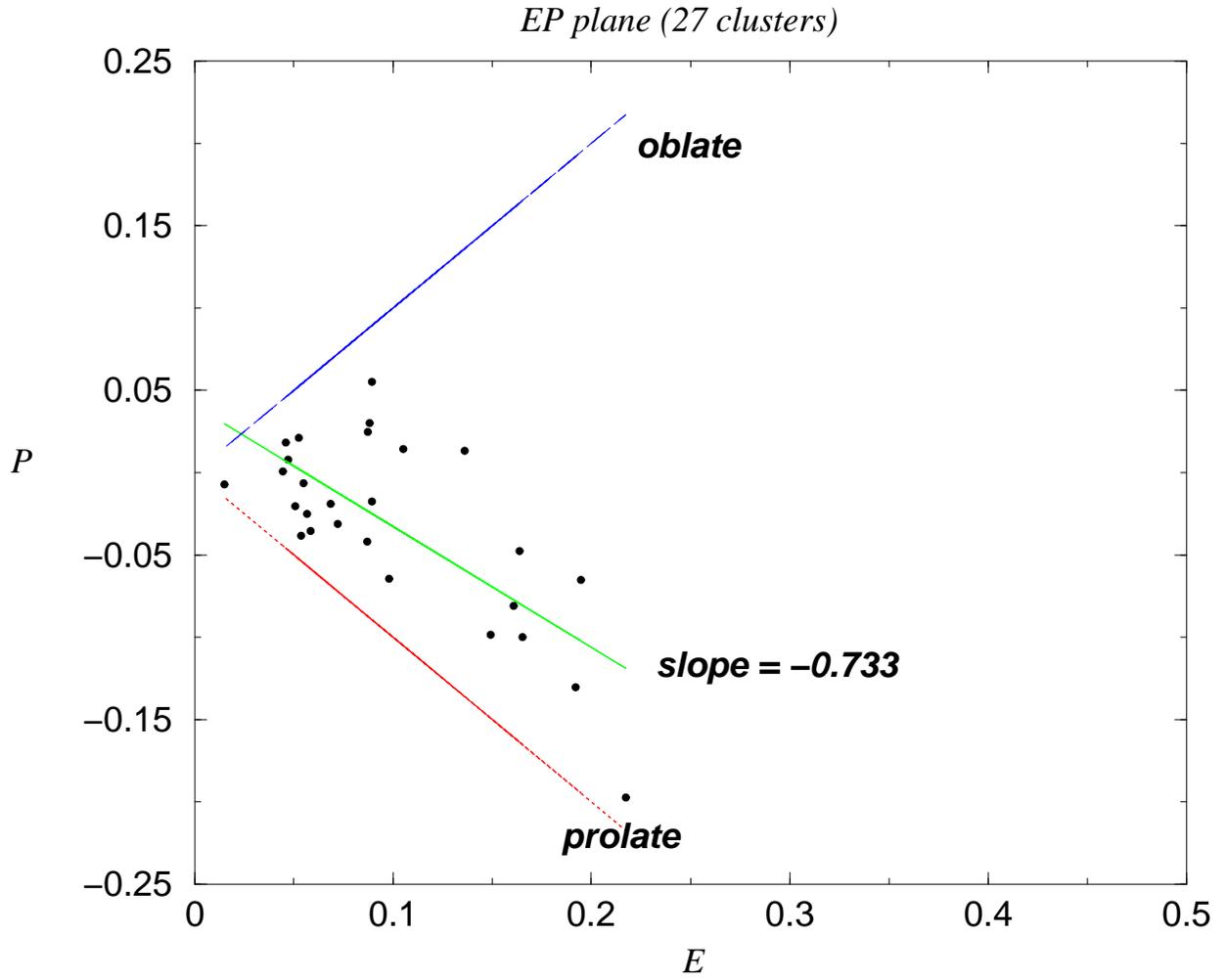}
\caption{\label{fig:ep_27.fig}  The morphology distribution of the original 27 clusters. The slope of a linear fit is -0.733.}
\end{figure}

\begin{figure}
\plotone{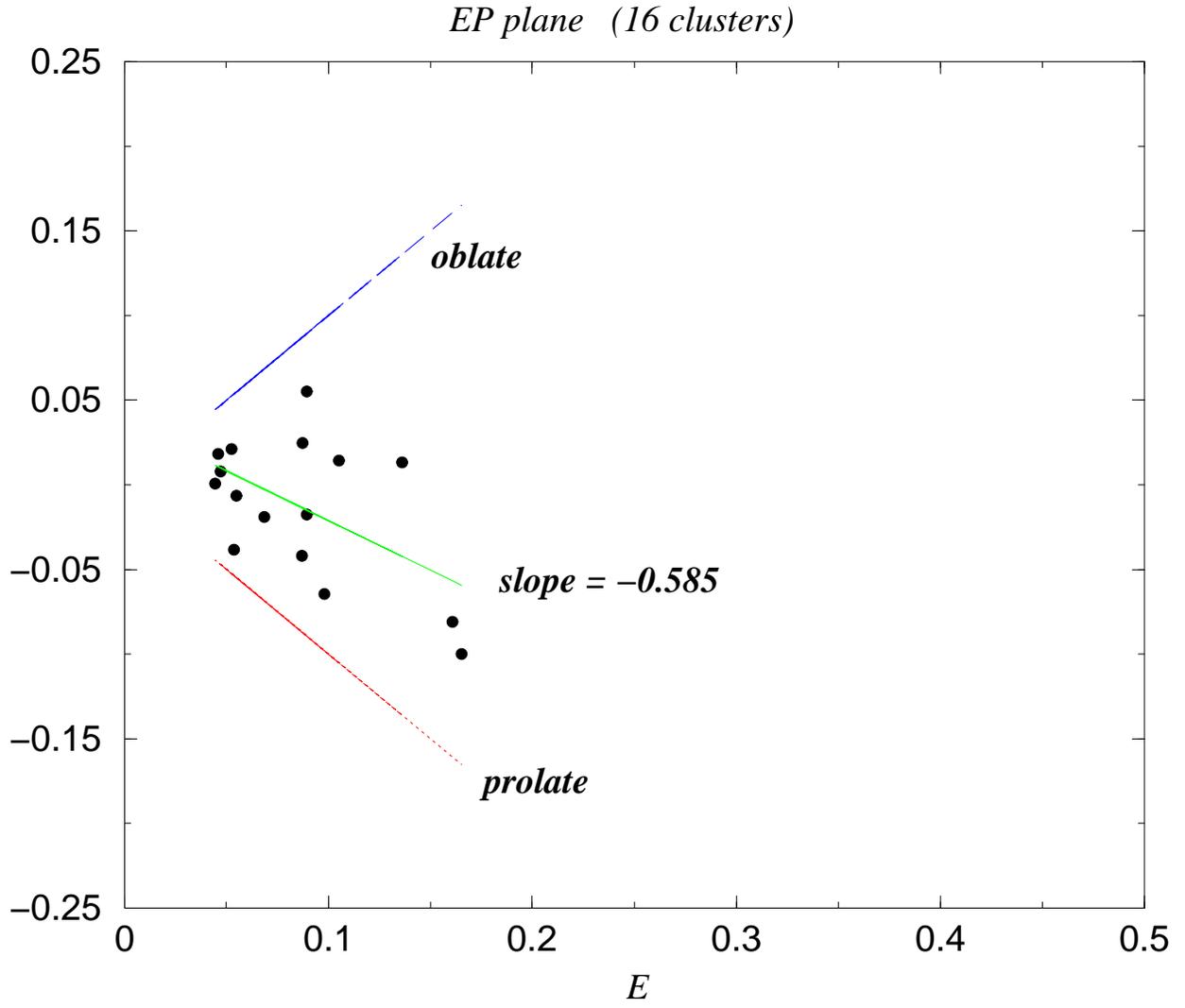}
\caption{\label{fig:ep_16.fig} The morphology distribution of the 16 clusters with well-fitted temperature profiles. The slope of a linear fit is -0.585.} 
\end{figure}

    In practice, clusters are aspherical while the $\beta$ model assumes that they are spherically symmetric. To describe the morphology of clusters in a more accurate way, we choose an elliptical model. An ellipsoid has three principal axes whose lengths we denote A, B and C from the longest one to the shortest one. The lengths of the three principal axes are propotional to the square root of the eigenvalues of the inertial tensor matrix.  We examine the oblateness or prolateness of clusters by calculating their E-P values (Thomas et al. 1998):
\begin{equation}
   E \equiv \frac{1}{2} \frac{B^2(A^2-C^2)}{B^2 C^2 + A^2 C^2 + A^2 B^2}
\end{equation}
\begin{equation}
   P \equiv \frac{1}{2} \frac{B^2 C^2 - 2A^2 C^2 + A^2 B^2}{B^2 C^2 + A^2 C^2 + A^2 B^2}
\end{equation}
   On the ellipticity-prolateness (E,P) plane, the strictly prolate clusters fall on the P = -E line while the strictly oblate clusters fall on the P = E line. Fig.~\ref{fig:ep_27.fig} shows the distribution of the original 27 clusters on the (E,P) plane with the slope of a linear fit equaling to -0.733. This indicates a strong prolateness among clusters. The prolateness of clusters is believed to be due to the cluster forming processes when they form from filaments. Similarly, Fig.~\ref{fig:ep_16.fig} displays the result of the 16 clusters with well-fitted temperature profiles (see $\S5$) and the slope from a linear fit of these profiles gives -0.585. Comparing the results from these two groups of clusters, the larger slope of the 16 clusters implies that they are less prolate than the original 27 cluters. Since relaxed clusters appear to be more spherical (less prolate), this result is consistent with the selection done by their temperature profiles.

\section{CLUMPING EFFECT}

Accretion shocks and merging effects during the history of cluster formation result in the small-scale density fluctuations of the intracluster medium (ICM). Therefore, all cluster atmospheres are expected to be clumpy. The clumpiness can be described by the clumping factor $C_n$, which is defined as $\frac{<n_e^2>}{<n_e>^2}$, where $n_e$ is the electron number density. Then the true angular diameter distance is (Birkinshaw 1999):
\begin{equation}
   D_A(true) \equiv C_n  D_A(estimated)
\end{equation}

\begin{figure}
\plotone{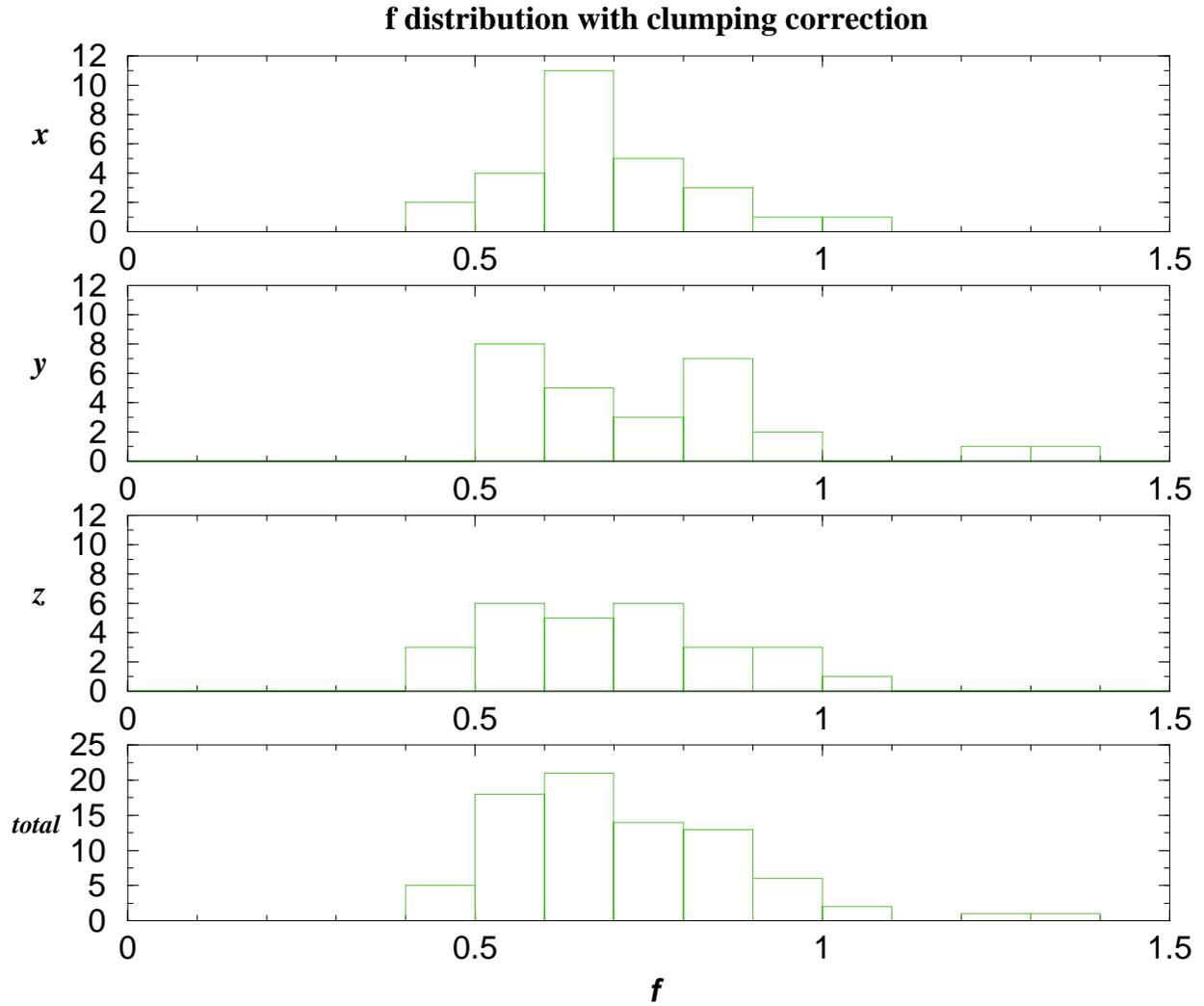}
\caption{\label{fig:f_clump.fig} Distributions of the normalized Hubble constant f with corrections of the clumping effect. The top three figures are the results of the original 27 clusters observing from x, y and z directions, respectively. The bottom one is the result by combining the data of the three lines of sight.}
\end{figure}

  This means that without accounting for the small-scale clumping, the value of $H_0$ based on Eqn.(\ref{eqn:Da.eqn}) is larger than the actual value. In our result, the values of $C_n$ range from 1.0 to 1.5 with a mean of 1.25 and a median of 1.17 (Table.~\ref{tbl:clumping.tbl}), which agrees with the result of Mathiesen $\&$ Evrard (1999). Applying the clumping correction lowers the mean and median values of f while the scattering of the f distribution decreases notably. Fig.~\ref{fig:f_clump.fig} shows the f distribution of the original 27 clusters after the clumping correction; Table.~\ref{tbl:clump_com.tbl} displays the results of the isothermal clusters before and after the clumping correction.

\section{CONCLUSION}
 Utilizing the combined analysis of SZE and X-ray data with a standard isothermal $\beta$ model to determine the value of $H_0$ results in the mean value of the normalized Hubble constant f being underestimated by 10 $\%$ and the median by 17 $\%$ compared to the actual value. There is a broad scattering in the f distribution and the standard deviation is around 32 $\%$ of the actual value if we parametrize the f distribution by a Gaussian. By setting the temperature of a whole cluster to be the emission-weighted temperature, the mean and median of f become 1.111 and 1.093, which is 10$\%$ higher than the actual value. However, this overestimate is offset by the clumping correction (Table.~\ref{tbl:clump_com.tbl}).

 The discovery of an universal temperature profile motivates us to propose a non-isothermal $\beta$ model to improve the analysis procedure. We refine our sample to 16 clusters which have well-fitted temperature profiles and apply the non-isothermal $\beta$ model to analyze them. Based on these 16 relaxed clusters, the mean value of f is underestimated by around 21 $\%$ if the isothermal $\beta$ model is applied and by around 12 $\%$ if the non-isothermal $\beta$ model is applied, which improves the mean value of f by 9 $\%$. By eliminating clusters with abnormal temperature profiles, we also eliminate non-relaxed clusters. The scattering in f distribution of these relaxed clusters is reduced to around 15 $\%$ of the actual value.

 We conclude that the broad scattering in the f distribution is due to the cluster morphology and the clumping effect. All our simulated clusters are at redshift 0.5 and a lot of them are dynamically young. We have shown the strong prolateness in the cluster morphology distribution of the original 27 clusters as well as the 16 refined clusters. The asphericity of clusters and random observational angles between the clusters' axes and observational line of sight result in the broad scattering in the f distribution. Another reason for the broad scattering is the clumpiness within the ICM. We have shown that with the correction of the clumping factor, the scattering in the f distribution decreases while the mean and median drop as well. 

 Therefore, in the real observation, we suggest to pick up relaxed clusters by their temperature profiles once the temperature structures can be measured. By applying a non-isothermal $\beta$ model to these clusters, we can derive the value of $H_0$. Under this method, we anticipate $\pm$ 15$\%$ scattering around the mean value of $H_0$ while the mean value is believed to be around 12$\%$ lower than the actual value.


\acknowledgments
 This work was supported by NSF grant AST-9803137 and NASA grant NAG5-7404.
All simulations were carried out on the SGI/CRAY Origin2000 at the NCSA.

\references
Birkinshaw, M., 1999, Phys. Rep., 310, 97 \\ 
Bliton, M. et al. 1998, MNRAS, 301, 609 \\
Bryan, G. L., Norman, M. L., Stone, J. M., Cen, R., \& Ostriker, J. P. 1995, Comput. Phys. Commun, 89, 149 \\
Bryan, G. L. \& Norman, M. L. 1998, \apj, 495, 80 \\
Bryan, G. L. 1999, Computing in Science and Engineering, 1:2, 46 \\
Cavaliere, A., Fusco-Femiano, R. 1976. A\&A, 49, 137
Cavaliere, A., Fusco-Femiano, R. 1978. A\&A, 70, 677
Donahue, Megan, 1996. \apj, 468, 79 \\
Freedman, W. L., et al. 1994, \apj, 427, 628 \\
Frenk, C. A. et al., 1999. \apj,525,554 \\
Gomez, P. L. et al. 1997, Astronomical Journal, 114(5), 1711 \\
Honda, H. et al. 1998, IAU, 188, 308 \\
Jones, C. \& Forman, W. 1984, \apj, 276, 818 \\ 
Loken, C. et al. 2002, \apj 579, 2, 571 \\
Markevitch, M., Forman, W.R., Sarazin, C.L., $\&$ Vikhlinin, A. 1998, \apj, 503, 77 \\
Mason, B. S. et al, \apj, 555, L11 \\
Mathiesen, B., \& Evrard, A. E., 1999. \apj, 520: L21-L24 \\
Mauskopf, P. A., 2000. \apj, 538, 505 \\
Metzler, Christopher A., \& Evrard, August E., 1994, \apj, 437, 564 \\
Myers, S. T., Baker, J. E., Readhead, A.C.S., \& Leitch, E.M., 1997. \apj, 485, 1 \\
Neumann, D. M., \& Bohringer, H. 1999, \apj, 512, 630 \\
Norman, M. L. $\&$ Bryan, G. L. 1999, in Numerical Astrophysics, eds. Miyama, S. M., $\&$ Hanawa, T., Kluwer Academic, p.19 \\
Norman, M. L., Daues, G., Nelson, E., Loken, C., Burns, J., Bryan, G. $\&$ Klypin, A. 1999, Large Scale Structure in the X-ray Universe, Proceedings of the 20-22 September 1999 Workshop, Aantorini, Greece, eds. Plionis, M. $\&$ Georantopoulos, I., Atlantisciences, Paris, France, p395 \\ 
Pierce, M., \& Tully, R. B. 1988, \apj, 330, 588 \\
Pierce, M., Welch, D., van den Bergh, S., McClure, R., Racine, R., \& Stetson, P. 1994, Nature, 371, 385 \\
Pointecouteau, E. et al. 2002, A$\&$A, 387, 56P \\
Reese, E. D. et al., 2000. \apj, 553,38 \\
Riess, A.B., Press, W.H., $\&$ Kirshner, R.P. 1995, \apj, 438, L17 \\
Roettiger, K. et al. 1995, \apj, 453, 634 \\
Roettiger,K., Stone, J., \& Mushotzky, Richard F., 1997. \apj, 482, 588 \\
Schaeffer, B. E. 1996, \apj, 460, L19 \\
Schmidt, B.P., Kirshner, R.P., Eastman, R.G., Phillips, M. M., Suntzeff, N. B., Hamuy, M., Maza, J., $\&$ Aviles, R. 1994, \apj 432,42
Sulkanen, M.E. 1999, \apj, 522, 59 \\
Sunyaev, R. A., \& Zeldovich, Ya. B. 1970, Comments Astrophys. Space Phys., 2, 66 \\
Sunyaev, R. A., \& Zeldovich, Ya. B. 1972, Comments Astrophys. Space Phys., 4, 173 \\
Sunyaev, R. A., \& Zeldovich, Ya. B. 1980, ARA\&A, 18, 537 \\  
Thomas, P.A. et al. 1998, MNRAS, 296, 1061 \\
Tonry, J. 1991, \apj, 373, L1 \\
van Dokkum, P. G., et al. 1999, \apj, 520, L95 
\begin{figure}
\plotone{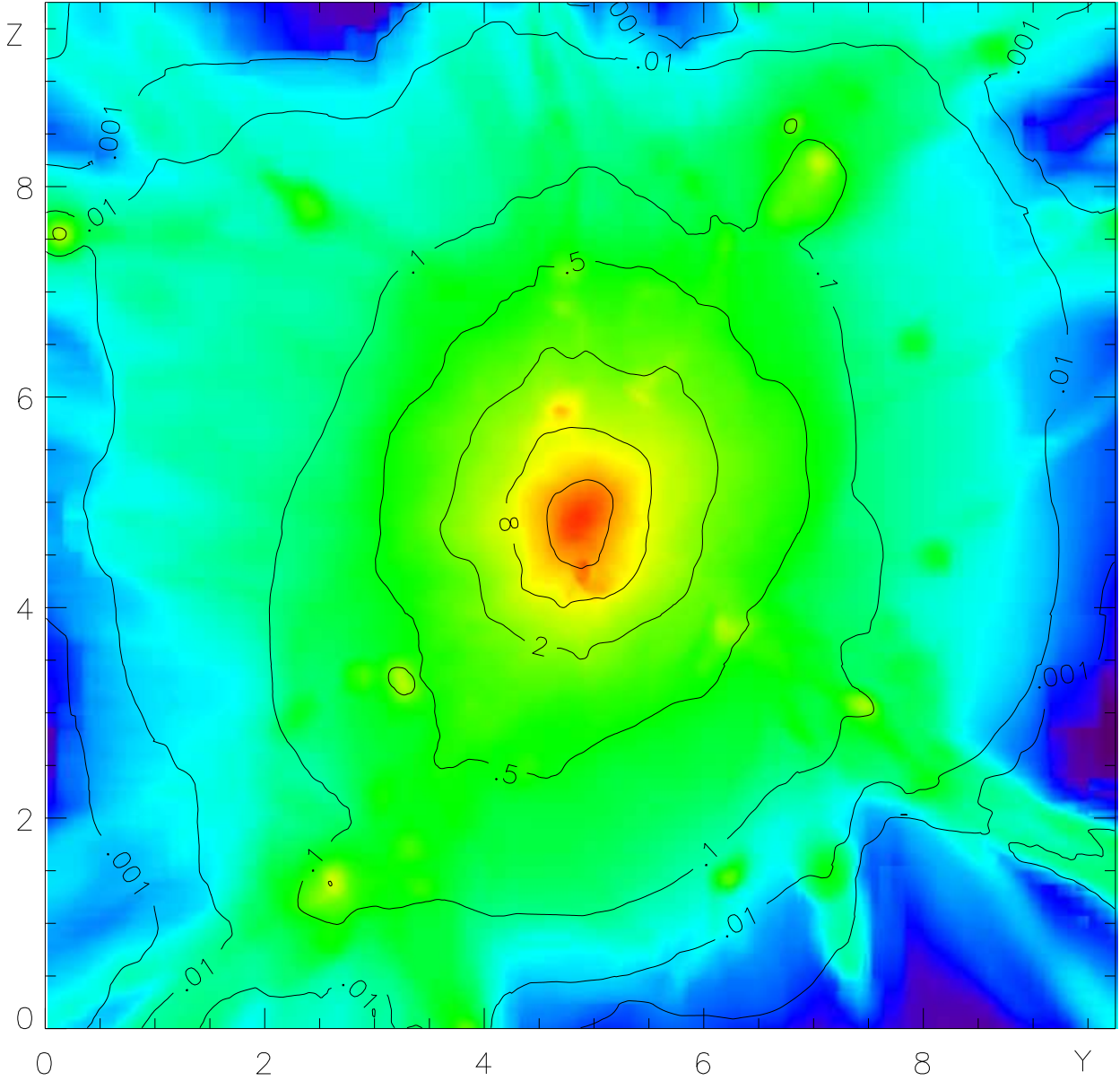}
\vspace{1 pt}
\caption{\label{fig:cluster} A 2D projection of our numerically simulated cluster. The color map is the X-ray surface brightness; the intensity drops from the central red core to the outer blue region. The contour map is the SZ y-parameter in units of $10^{-4}$.}
\end{figure}
\begin{figure}
\plotone{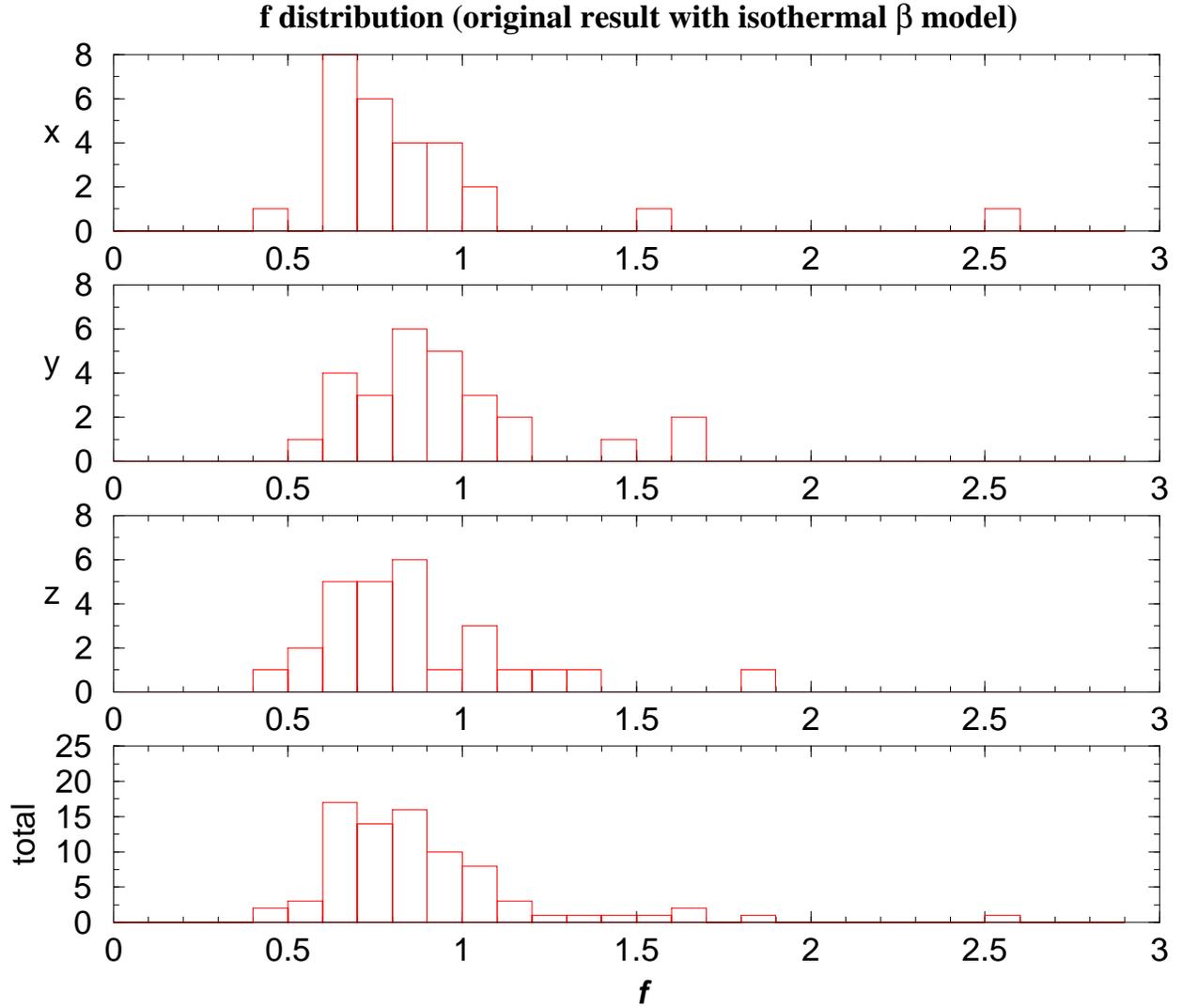}
\caption{\label{fig:f_noclump.fig} The distributions of the normalized Hubble constant f derived from the original 27 clusters using the isothermal $\beta$ model. The top three figures display the results observing along the x, y, and z directions, respectively; the bottom figure shows the result combining the three lines of sight data.}
\end{figure}



\clearpage
\begin{table*}
\begin{center}
\caption{The history of the measurements of $H_0$}
\vspace{1cm}
\label{tbl:H_0.tbl}
\begin{tabular}{|l|c|c|c|}
\tableline\tableline
year    & value        & Method                   &   Reference    \\
\hline
1988    &  85 $\pm$ 10 & Tully-Fisher relation    &   Pierce \& Tully \\ 
\hline
1994    &  73 $\pm$ 7  & Type II supernovae and   & Schmide et al. \\
        &              & the expanding photoshpere method &           \\
\hline
1994    &  80 $\pm$ 17 & Cepheid variable stars in the Virgo Cluster & Freedman et al. \\
        &              & with space-based measurement                &                 \\  
\hline
1994    &  87 $\pm$ 7 & Cepheid variable stars in the Virgo Cluster & Pierce et al. \\
        &              & with ground-based measurement                &                 \\
\hline
1995    &  67 $\pm$ 7 & Type Ia - based value  &   Riess, Press, $\&$ Kirshner \\
\hline
1996    &  55 $\pm$ 3 & Type Ia supernovae    &   Schaefer \\
\hline

\end{tabular}
\end{center}
\end{table*}

\clearpage
\begin{table*}
\begin{center}
\caption{The mean, median and standard deviation of the normalized Hubble constant for the original 27 clusters. $f_x$, $f_y$ and $f_z$ denote the results observing from the x, y and z directions, respectively.\label{tbl:f_dis_ori.tbl}}
\vspace{0.5cm}
\begin{tabular}{|l|cccc|}
\tableline
$f$     & $f_{x}$ & $f_{y}$  &  $f_{z}$  &  $f_{all}$    \\
\hline
mean    &  0.876    &   0.939   &    0.876    &   0.897  \\
median  &  0.763    &   0.891   &    0.8      &   0.829  \\
sig     &  0.393    &   0.274   &    0.287    &   0.32   \\
\tableline
\end{tabular}
\end{center}
\end{table*}
\begin{table*}
\begin{center}
\caption{The mean, median and standard deviation of the normalized Hubble constant for the 27 isothermal clusters. Here $f_x$, $f_y$ and $f_z$ denote the results observing from the x, y and z directions, respectively.\label{tbl:f_dis_iso.tbl}}
\vspace{0.5cm}
\begin{tabular}{|l|cccc|}
\tableline
$f$     & $f_{x}$  &    $f_{y}$  &    $f_{z}$  &    $f_{all}$  \\
\hline
mean    &  1.051   &   1.178     &    1.104    &    1.111      \\
median  &  0.966   &   1.193     &    1.119    &    1.093      \\
sig     &  0.303   &   0.278     &    0.306    &    0.297      \\
\tableline
\end{tabular}
\end{center}
\end{table*}

\begin{table*}
\begin{center}
\caption{The mean, median, and standard deviation of $\beta$ value distribution. $\beta_x$, $\beta_y$, and $\beta_z$ are the results from the x, y, and z observational lines of sight. $\beta_{all}$ is the result combining the three lines of sight data.\label{tbl:beta.tbl}}
\vspace{0.5cm}
\begin{tabular}{|l|cccc|}

\tableline
$\beta$ & $\beta_{x}$ & $\beta_{y}$ & $\beta_{z}$ & $\beta_{all}$ \\
\tableline
mean    &  0.621      &    0.605    &    0.613    & 0.613         \\
median  &  0.613      &    0.597    &    0.599    & 0.599         \\
sig     &  0.144      &    0.134    &    0.146    & 0.140         \\
\tableline
\end{tabular}
\end{center}
\end{table*}
\begin{table*}
\begin{center}
\caption{Comparing the results of the 16 clusters with well-fitted temperature profiles using the isothermal $\beta$ model and the non-isothermal $\beta$ model.\label{tbl:Bmodel_compare.tbl}}
\vspace{0.5cm}
\begin{tabular}{|c|c|c|}
\tableline
model   & isothermal $\beta$ model & non-isothermal $\beta$ model \\
$f$     & $f$      &    $f$   \\
\tableline
mean    &  0.804   &    0.883     \\
median  &  0.796   &    0.884     \\
sig     &  0.150   &    0.152     \\
\tableline
\end{tabular}
\end{center}
\end{table*}
\begin{table*}
\begin{center}
\caption{The mean, median, and standard deviation of the clumping factor distribution for the original 27 simulated clusters.\label{tbl:clumping.tbl}}
\vspace{0.5cm}
\begin{tabular}{|l|c|}
\tableline
    &        clumping factor        \\
\tableline
mean    &  1.254   \\
median  &  1.168   \\
sig     &  2.929   \\
\tableline
\end{tabular}
\end{center}
\end{table*}

\begin{table*}
\begin{center}
\caption{Comparing the results of the 27 isothermal clusters before and after clumping correction using isothermal $\beta$ model.\label{tbl:clump_com.tbl}}
\vspace{0.5cm}
\begin{tabular}{|c|c|c|}
\tableline
model   & before clumping correction & after clumping correction \\
$f$     & $f$      &    $f$   \\
\tableline
mean    &  1.111   &    0.91     \\
median  &  1.093   &    0.93     \\
sig     &  0.297   &    0.228    \\
\tableline
\end{tabular}
\end{center}
\end{table*}

\end{document}